# Tunable Indirect-Direct Transition of Few-Layer SnSe via Interface Engineering

Hansika I. Sirikumara and Thushari Jayasekera*
*Department of Physics, Southern Illinois University, Carbondale*
(Dated: March 13, 2017)

Tin Selenide (SnSe) is one of the best thermoelectric materials reported to date. The possibility of growing few-layer SnSe helped boost the interest in this long-known, earth abundant material. Pristine SnSe in bulk, mono- and few-layer forms are reported to have indirect electronic bandgaps. Possible indirect-direct transition in SnSe is attractive for its optoelectronic-related applications. Based on the results from first principles Density Functional Theory (DFT) calculations, we carefully analyzed electronic band structures of bulk, and bilayer SnSe with various interlayer stackings. We report the possible stacking-dependent indirect-direct transition of bilayer SnSe. By further analysis, our results reveal that it is the directionality of interlayer interactions that determine the critical features of their electronic band structures. In fact, by engineering the interface stacking between layers, it is possible to achieve few-layer SnSe with direct electronic band gap. This study provides fundamental insights to design few-layer SnSe and SnSe heterostructures for electronic/optoelectronic applications, where the interface geometry plays a fundamental role in device performance.

The discovery of single layer graphene and transition metal dichalcogenides have sparked a series of high profile discoveries that impact numerous electronic and optoelectronic applications[1–7]. Another addition to the 2D family is layered semiconductors such as group IV-monochalcogenides, also referred to as four-six-enes, out of which tin selenide (SnSe) is the focus of the current investigation[8–11]. Layered structures of these four-six-enes are mediated through the displacement of adjacent layers to minimize the non-bonding interaction of electron lone pairs on cation sites[12]. The electron lone pair repels the neighboring atoms, thus creating a van der Waals (vdW) gap between chemically bound layers[13]. Four-six-enes with vdW gap are ideal for two-dimensional electron dynamics.

Tin selenide has been of interest for over decades for various applications such as photovoltaics[14–16]. Recently reported high thermoelectric figure of merit of earth-abundant SnSe has immensely renewed the interest in the family of group IV-monochalcogenides[17–22]. The possibility of growing mono to few-layers of selenide-based compounds[23,24] has also helped boost the interest in these long-known materials[25,26]. Their atomic structure derived from distorted rocksalt structure is highly anisotropic. SnSe can be thought of as a binary counterpart of phosphorene, in which the direct electronic band gap overlaps with the visible spectrum[27]. Structural properties of SnSe are unique. Unlike other 2D materials, few-layer SnSe shows strong interlayer coupling through ultralow interlayer breathing modes. Recent Raman frequency measurements support this unusual bonding[28]. It is also reported that monolayer phosphorene, which is isostructural to SnSe cannot be isolated via mechanical exfoliation[29], which also could be due to strong interlayer coupling. Our investigation finds that the nature of interlayer interactions has a significant role in determining the critical features of electronic band gap in few-layer SnSe.

All group IV-monochalcogenides are reported to be indirect bandgap semiconductors[10,20,30], while a direct bandgap is preferred for some applications such as optoelectronics. There has been a considerable interest in exploring for ways to achieve direct bandgap group IV-monochalcogenides. C. Kamal *et al.* reported that even though pristine-bulk group IV-monochalcogenides are reported to have indirect bandgap, the difference between direct and indirect electronic bandgaps are relatively small, such that an external influence can possibly result in a indirect-direct transition in these materials. In fact, C. Kamal *et al.* suggested that mechanical strain is a possible path for inducing indirect-direct transition in monolayer group IV-monochalcogenides[31].

The goal of this work is to achieve a deeper understanding of the electronic band structure of group IV-monochalcogenides at the atomic level. Such understanding offers fundamental insights, which are expected to result in novel avenues for further engineering of these materials; especially for reverse engineering - to design atomic configurations for desired electronic properties. This approach is even more promising, when advanced crystal growth techniques are reaching the capability of synthesizing pre-designed crystal structures. In this work, we have identified that the strong interlayer coupling of SnSe determines the critical features of its electronic band structure.

Crystal structure of SnSe is orthorhombic, a black phosphorous analogue, where atoms are arranged in two-atom thick layers with zigzag and armchair patterns in $\hat{a}$ and $\hat{b}$ directions[17,20,30,32,33]. These two-atom thick layers are stacked in the $\hat{c}$ direction with vdW interactions. Fig.1 (a) depicts the atomic structure of SnSe. This structure can also be thought of as a distorted rocksalt structure. For a comparison, Fig.1 (b) depicts the atomic structure of undistorted rocksalt structure, in which each $A$ atom is chemically bound to six $B$ atoms with octahedral geometry. In the perfect rocksalt structure all $A$-$B$-$A$ angles are $90^0$. In SnSe, each two-atom thick layer is distorted from the perfect rocksalt type, deviating the Sn-Se-Sn angle from $90^0$. There are strong chemical bonds within two-atom thick layers, which are bound together with vdW interactions along the direction perpendicular to the plane. Translational periodicity in zigzag, armchair and out of planar directions for the optimized atomic configuration are $4.22^0 A$, $4.36^0 A$, and $11.86^0 A$.

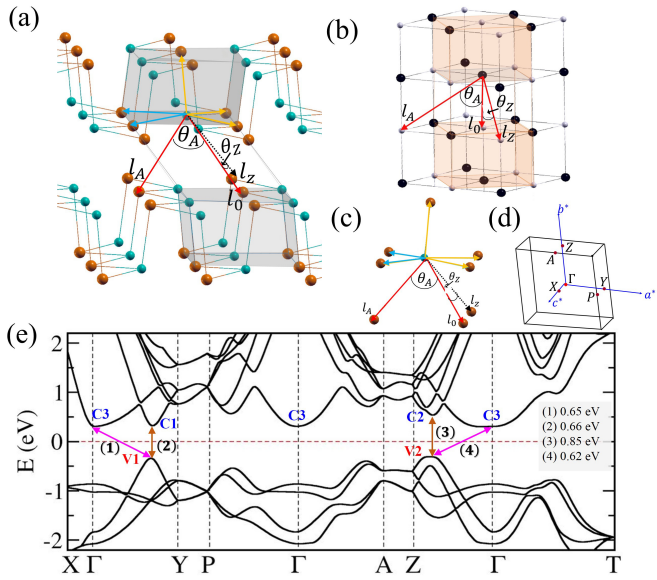

FIG. 1. (a) Atomic structure, (c) Closer view of Se-coordination, (d) Sketch of the first Brillouin Zone, and (e) the electronic structure of Bulk SnSe. Atomic structure of undistorted rocksalt structure is shown in panel (b) for a comparison.

Fig.1 (c) clearly shows the coordination of a Se atom. A given Se atom is covalently bound to 3 Sn atoms (marked with yellow solid lines in Fig. 1(a) and (c)) and it has 4 longer bonds with Sn atoms (marked with blue and red solid lines in the Fig. 1(a) and (c)). Two of the second nearest Sn neighbors for a given Se atom lie within the same layer (those marked with blue solid lines). The other two second nearest neighbor Sn atoms lie in the adjacent layer (those marked in red solid lines). Our investigation reveals that the directionality of these interlayer second nearest neighbors (those marked with red solid lines) plays a major role in determining the critical features of the electronic band structure of SnSe. The direction of the interlayer second nearest neighbors is determined by the degree of distortion of the atomic configuration from rocksalt structure. Our main focus is on the interlayer configuration, while each two-atom thick layer is distorted from rocksalt configuration. The two-atom thick layers are shaded in both distorted and undistorted structures in Fig.1. We quantify the degree of interlayer distortion by two angles $\theta_A$ and $\theta_Z$ as shown in Fig.1, which are both equal to ArcTan $[\sqrt{2}]$ ($54.7^0$) in rocksalt structure (Fig.1 (b)). Directionality of interlayer interactions is related to the changes in $\theta_A$ and $\theta_Z$. To ease the explanation, we name the closest interlayer interatomic distance by $l_0$ (which is usually $\sim 3.4-3.8^0 A$), the closest interatomic distance in the armchair and zigzag directions by $l_A$ and $l_Z$. In the case of bulk SnSe structure, $\theta_A = 69^0$ and $\theta_Z = 49^0$ (Fig 1-(a)), where $\theta_Z$ is decreased and $\theta_A$ is increased from the octahedral coordination of undistorted rocksalt structure. As $\theta_Z$ increases, $l_Z$ decreases. On the other hand, $l_A$ increases as $\theta_A$ decreases. For instance, in the case of bulk SnSe shown in the Fig.1, $\theta_A = 69^0$ and $\theta_Z = 49^0$, $l_A < l_Z$, i.e. interlayer interactions extend in the armchair direction, which we named as $AB_A$-type interactions in the rest of the text. This fact is further visualized in Fig.5. Our aim is to relate the structural changes in $\theta_A$, $\theta_Z$, $l_A$ and $l_Z$ to the changes in electronic band structure of SnSe.

Electronic structure of bulk SnSe is shown in Fig.1 (e), which is in agreement with already published results[17,34–36]. Conduction band minimum (CBM) is at the $\Gamma$ point, which is marked as $C_3$. We have also identified two other local CBM's: $C_1 \equiv (0.35, 0.0, 0.0)$, along $\Gamma \to Y$ direction and $C_2 \equiv (0.0, 0.38, 0.0))$, along $\Gamma \to Z$ direction. Valance band maximum (VBM) is along $\Gamma \to Y$ direction ($V_1 \equiv (0.35, 0.0, 0.0)$) and there is another local VBM along $\Gamma \to Z$ direction ($V_2 \equiv (0.0, 0.38, 0.0)$). Electronic band structure implies that bulk SnSe has an indirect electronic bandgap of 0.62 eV for the $V_2 \to C_3$ transition. The energy associated with the direct transition, $V_1 \to C_1$ (0.66 eV) is also close to the energy of the indirect gap $V_2 \to C_3$, which suggests a possibility of indirect-direct transition via external influence to the system[31].

In order to achieve a deeper understanding about the relation between the interlayer atomic configuration and electronic bands, we have investigated the electronic band structure of artificially created SnSe systems with various interlayer stackings. All calculations were done using first principles Density Functional Theory (DFT) as it is implemented in the *Quantum Espresso* package[37]. At least 15 $^0 A$ vacuum space was used to separate the structure from its periodic image. Projector augmented wave (PAW) scheme, and the Generalized gradient approximation of Perdew, Burke and Ernzerhof (GGA-PBE) were used for the exchange and correlation functional with a 30 Ry energy cutoff for plane wave expansion[38]. A $12 \times 12 \times 1$ Monkhorst-Pack grid was used to sample the Brillouin zone. All geometries were optimized to forces less than 0.025 eV/$^0 A$.

We considered a structure, in which one of the two-atom thick layer is shifted through half of the lattice constant $a$ along the zigzag direction (Fig.2). This structure is defined by $\theta_A = 50^0$ and $\theta_Z = 66^0$ as marked in Fig.2(a). For this structure, $l_Z < l_A$, i.e. directionality of interlayer interaction extends in the zigzag direction (which we name as $AB_Z$-type interactions). Electronic bandgap remains indirect with a 0.57 eV associated with $V_2 \to C_3$ transition. One important thing to notice is that, the energy of the $C_2$ point ($E_{C2}$) is relatively lower than $E_{C1}$, yet $C_3$ (at the $\Gamma$ point) remains the CBM in this new structure.

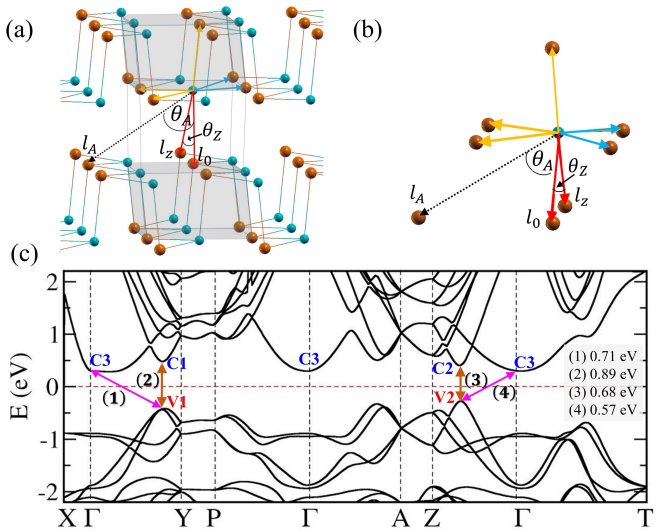

FIG. 2. (a) Atomic structure, (b) Closer view of Se-coordination, and (c) the electronic structure of artificially designed bulk SnSe with $AB_Z$-type interlayer interactions.

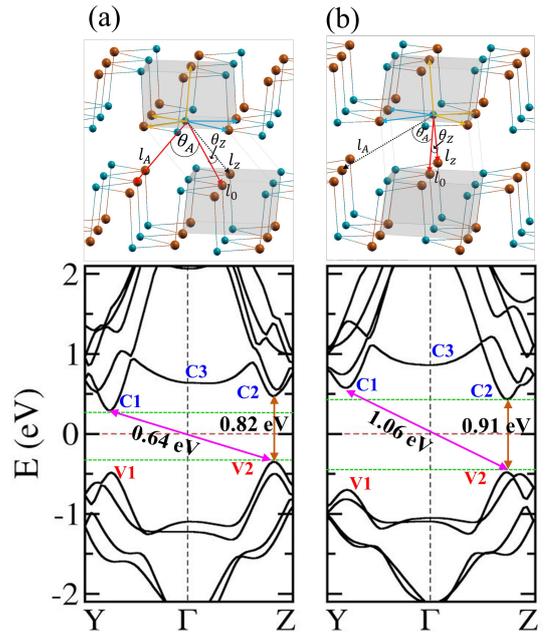

FIG. 3. Atomic and electronic band structure of bilayer SnSe with (a) $AB_A$ and (b) $AB_Z$-type interlayer interactions, which show indirect and direct electronic bandgap respectively.

Structural parmeters for two bulk-SnSe configurations considered in Fig.1 and Fig.2 are summarized in Table 1. Noting that the VBM stays at $V_2$ point and stacking-dependent changes in $E_{C1}$ and $E_{C2}$, we investigated the electronic band structures of bilayer SnSe for the above two configurations with $AB_A$ and $AB_Z$ type interlayer interactions.

TABLE I. Structural details for all bilayer SnSe configurations considered in Fig. 1 and Fig.2: closest interatomic distances $l_A$, $l_0$ and $l_Z$ and angles, $\theta_A$ and $\theta_Z$

| Structure | $l_A$ | $l_0$ | $l_Z$ | $\theta_A$ | $\theta_Z$ |
|---|---|---|---|---|---|
| Fig. 1 | 4.03 | 3.76 | 5.66 | 69 | 49 |
| Fig. 2 | 5.62 | 3.90 | 3.91 | 50 | 66 |

Interestingly, direct and indirect electronic bandgaps were found for the bilayer SnSe with $AB_Z$ and $AB_A$-type interlayer interactions. The VBM is located at $V_2$ in both the structures. One of the most significant changes in the electronic band structures of the bilayers compared with those of bulk is that $E_{C3}$ (i.e $\Gamma$ point) increases with relative to $E_{C1}$ and $E_{C2}$, i.e. CBM is no longer at the $C_3$ point. Another important difference between the electronic band structures of the two systems is that $E_{C2}$ is lower than $E_{C1}$ for the system with $AB_Z$-type interactions, where as it is opposite for the system with $AB_A$-type interactions. This implies that $AB_Z$ bilayer SnSe is a direct bandgap semiconductor. The nature of the bandgap has a direct influence from the interlayer interactions in bilayer-SnSe. When $\theta_A < \theta_Z$, interlayer interactions extend in zigzag direction (i.e. $l_Z < l_A$), which results in a direct gap. For both the bilayer systems we considered, VBM lies at the $V_2$ point. Relative energy difference at the $C_1$ and $C_2$ determines the nature of the bandgap (if it is a indirect or direct) for the bilayer systems. Stacking dependent indirect-direct transition is not observed in the bulk system because the CBM lies at the $C_3$ (i.e. $\Gamma$) point.

It is hypothesized that, directionality of interlayer interactions of SnSe, which can be described using the degree of distortion from rocksalt structure, determines the direct-indirect nature of bilayer SnSe. Fig.4 shows two other stacking arrangements, which are resulted from (a) both layers are directly stacked on top of each other ($\theta_A = 55^0$ and $\theta_Z = 51^0$), and (b) one of the layer is shifted through half of the lattice constant $b$ along the armchair direction ($\theta_A = 71^0$ and $\theta_Z = 49^0$). In both the cases, VBM is found at the $V_2$, and $E_{C1}$ is lower than that of $E_{C2}$ (i.e. CBM lies at $C_1$), thus an indirect electronic bandgaps result in both cases. It is also important to note that $\theta_A > \theta_Z$ and $l_A < l_Z$ for both structures, i.e. interlayer second nearest neighbor interactions extend in the armchair direction as they are marked in the Fig.4.

TABLE II. Structural details for all bilayer SnSe configurations considered in Fig. 3 and Fig.4: closest interatomic distances $l_A$, $l_0$ and $l_Z$ and angles, $\theta_A$ and $\theta_Z$.

| Structure | $l_A$ | $l_0$ | $l_Z$ | $\theta_A$ | $\theta_Z$ |
|---|---|---|---|---|---|
| Fig. 3 (a) | 3.90 | 3.68 | 5.61 | 69 | 49 |
| Fig. 3 (b) | 5.56 | 3.77 | 3.76 | 52 | 69 |
| Fig. 4 (a) | 5.22 | 3.40 | 5.48 | 55 | 51 |
| Fig. 4 (b) | 3.92 | 3.61 | 5.58 | 71 | 49 |

Structural details for the SnSe bilayer configurations considered in Fig.3 and Fig.4 are summarized in Table 2. Out of all bilayer structures we considered (Fig.3 and Fig.4), the one shown in the Fig. 3 (b) with $AB_Z$-type interlayer interactions shows a direct electron band gap with $V_2 \to C_2$ transition. All other structures has $AB_A$-type interlayer interactions and have an indirect electron bandgap with $V_2 \to C_1$ transition. The former structure with direct band gap can be thought of as two SnSe layers stacked together with one layer shifted with relative to the other through half of the lattice constant $a$ in zigzag direction.

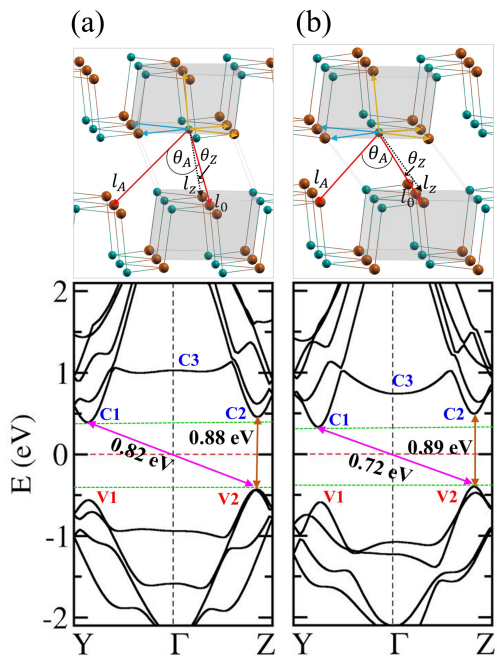

FIG. 4. Atomic configuration and electronic band structure of bilayer SnSe with (a) both layers are directly stacked on top of each other, and (b) one of the layer is shifted through half of the lattice constant $b$ along the armchair direction. Note that both structures have $AB_A$ interactions and indirect electron bandgaps.

To further understand the relation between indirect-direct transition and interlayer interactions, we have analyzed electronic band structure for 20 relative shifts ($\Delta$) between the two layers along the zigzag direction and the results are shown in Fig.5. In this graph, $\Delta = 0$ represents the structure shown in Fig. 4 (a), where second layer is on top of the first layer, whereas $\Delta = 0.5$ is the structure shown in Fig. 3 (b). As it was pointed out earlier, relative difference of $E_{C1}$ and $E_{C2}$ determines the nature of the electronic band gap. The variation of $E_{C1}$, $E_{C2}$, and $E_{V2}$ as a function of the relative shift, $\Delta$ is shown in Fig. 5 (a). At $\Delta = 0$, $E_{C1} < E_{C2}$ that $V_2 \to C_1$ transition is more energetically favorable than $V_2 \to C_2$ transition, which implies an indirect bandgap. This trend changes at $\Delta = 0.14$ (which is marked by a vertical dashed-red line in Fig.5 (a)) resulting $E_{C2} < E_{C1}$, that is CBM turns from $C_1$ to $C_2$. This implies a direct electron bandgap associated with $V_2 \to C_2$ transition.

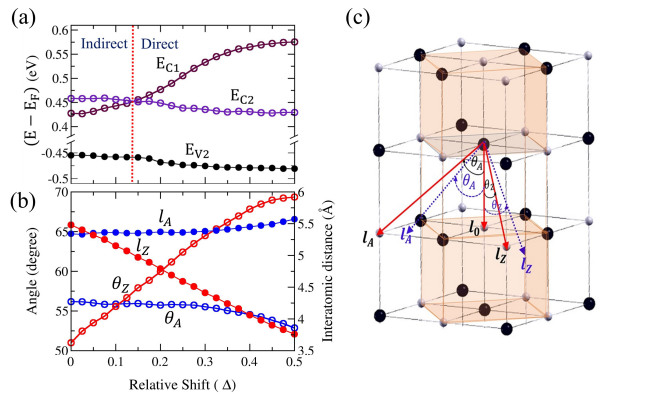

FIG. 5. (a) The variation of $E_{C1}$, $E_{C2}$ and $E_{C3}$, (b) The variation of $\theta_A$ (blue empty spheres), $\theta_Z$ (red empty spheres), $l_A$ (blue solid spheres) and $l_Z$ (red solid spheres) as a function of the relative shift between two layers of SnSe. Panel (c) shows (based on the rocksalt structure) the fact that as $\theta_z$ increases, $l_z$ decreases.

As we have hypothesized, indrect-direct trnasition of bilayer SnSe correlates to the directionality of interlayer interactions, which can be explained using the two angles $\theta_A$ and $\theta_Z$. The sketch in the Fig.5 (b) shows fact that $l_Z$ decreases as $\theta_Z$ increases, and $l_A$ increases as $\theta_A$ decreases. Fig. 5 (c) visualizes this fact based on the undistorted rocksalt structure. When the interlayer distortions are such that $\theta_A < \theta_Z$, the interlayer second nearest neighbor interactions extend in the zigzag direction, which then result in a direct-gap bilayer SnSe. The changes do not happen at the exact critical point as shown in the Fig. 5 (a) and (b). However, we clearly see a correlation in these three facts; changes in $\theta_A/\theta_Z$, changes in $l_A/l_Z$ and indirect-direct transition. It is also important to note that, when there are more than bilayers, few-layer SnSe shows a direct bandgap as long as all the interlayer interactions are $AB_Z$ type.

In summary, based on the results from first principles DFT calculations, we showed that the interlayer interactions plays a major role in the direct-indirect transition of few-layer SnSe. There is a possibility of achieving direct band gap few-layer SnSe by modifying their atomic configuration at the interface. This study provides fundamental insights for designing few-layer SnSe and SnSe heterostructures for device applications, where the interface geometry plays a fundamental role in device performance.

Authors would like to thank the financial support through a SEED grant from Materials Technology Center at Southern Illinois University, and the computer facilities provided by Southern Illinois University which were partly funded by National Science Foundation Awards 0721623 and 1032778.